\documentclass[twocolumn,showpacs,amsmath,amstex,amssymb,mathfonts,prl]{revtex4-1}
\usepackage{graphicx}

\usepackage{tipa}
\usepackage{bm}

\usepackage{amsmath}
\usepackage{amssymb}
\usepackage{amsthm}
\usepackage{amsfonts}
\usepackage{float}

\begin{document}

\title{An Effective Model for Traffic Dynamics and the Nature of the Congested Phase}

\author{Bo Yang and Christopher Monterola}
\affiliation{Complex Systems Group, Institute of High Performance Computing, A*STAR, Singapore, 138632.}
\date{\today}

\date{\today}
\begin{abstract}
A simple algorithm for constructing an effective traffic model is presented. The algorithm uses statistically well-defined quantities extracted from the flow-density plot, and the resulting effective model naturally captures and predicts many quantitative and qualitative empirical features of the highway traffic, especially with the presence of an on-ramp bottleneck. The simplicity of the effective model provides strong evidence that stochasticity, diversity of vehicle types and modeling of complicated driving behaviors are \emph{not} fundamental to many observations in the complex real traffic dynamics. We also propose the nature of the congested phase can be well characterized by the long lasting transient states of the effective model, from which the wide moving jams evolve.

\end{abstract}

\maketitle 

Modeling the dynamics of highway traffic flow has been the endeavor of researchers in many disciplines for the last fifty years\cite{dogbe, as, helbing, kernerbook}. Various different models have been proposed to describe both the free and the congested phase of the traffic flow\cite{bando,JiangR_PRE01,PengGH_PhyA13,GLW_PhyA08,kerner1, kerner2,stepan, nagel,kim,naka,xue,nishinari}. In general most of these models can describe the low density free flow phase, and the wide moving jams when the density is high. Kerner\cite{kernercrit} first suggested that some essential empirical features are not captured by most of these models; there exists a ``synchronized phase" that can be distinguished by a scattering of data points covering a two-dimensional region on the flow-density plane. This phase is qualitatively different from the wide moving jams, particularly when a bottleneck at the highway is present\cite{kernerexp1,kernerexp2}. This raises the questions of the relevance of the popular general motor (GM) model classes to real traffic systems, because these models only describe a two-phase transition\cite{kernercrit}. 

Most of the three-phase models are constructed by putting in a ``synchronization gap" by hand\cite{kerner1,kerner2,kerner3}, at the cost of making the models more sophisticated with significantly more parameters. These, together with other three-phase models\cite{kimca}, reproduce the ``synchronized phase" with a multitude of steady states in the congested phase. However, Helbing et.al\cite{helbing1,helbing2} pointed out that the characterization of the complex congested states of the traffic flow as a single synchronized phase is delicate, and with properly adjusted parameters some GM models \emph{can} reproduce many empirical observations at the highway bottleneck. One should also note it is not well understood if the empirical data in the congested phase comes from equilibrium/steady traffic conditions, or from slowly evolving transient ones. Inhomogeneous road conditions and vehicle types\cite{helbing3}, as well as stochastic driving behaviors can also contribute to the scattering of the flow-density plot. It is thus important to understand how complex a model needs to be to capture the essential features of the empirical data.

In this Letter, we attempt to address this problem by presenting a simple algorithm in constructing an effective microscopic traffic model based on the macroscopic empirical data. Instead of trying to model the microscopic driving behaviors of individual drivers, we only use statistically robust quantities from the flow-density plot. We also adopt a reductionist point of view by looking at the simplest model possible, in order to strip away all unnecessary elements that are not fundamental in capturing the observed empirical features. For this reason our deterministic effective model is based on a modified optimal velocity function with identical drivers on a homogeneous road; more importantly, the model shows simple driving behaviors without speed and environmental adaptations are sufficient in reproducing many empirical features numerically. In addition, the model shed light on the nature of the traffic congestion and provide evidences they can be described by long lasting transient states. The wide moving jams emerge from these transient states, thus they are in most cases observed to evolve from the congested traffic instead of the free flow traffic\cite{kernerbook}.

\begin{figure}
\includegraphics[width=6.8cm]{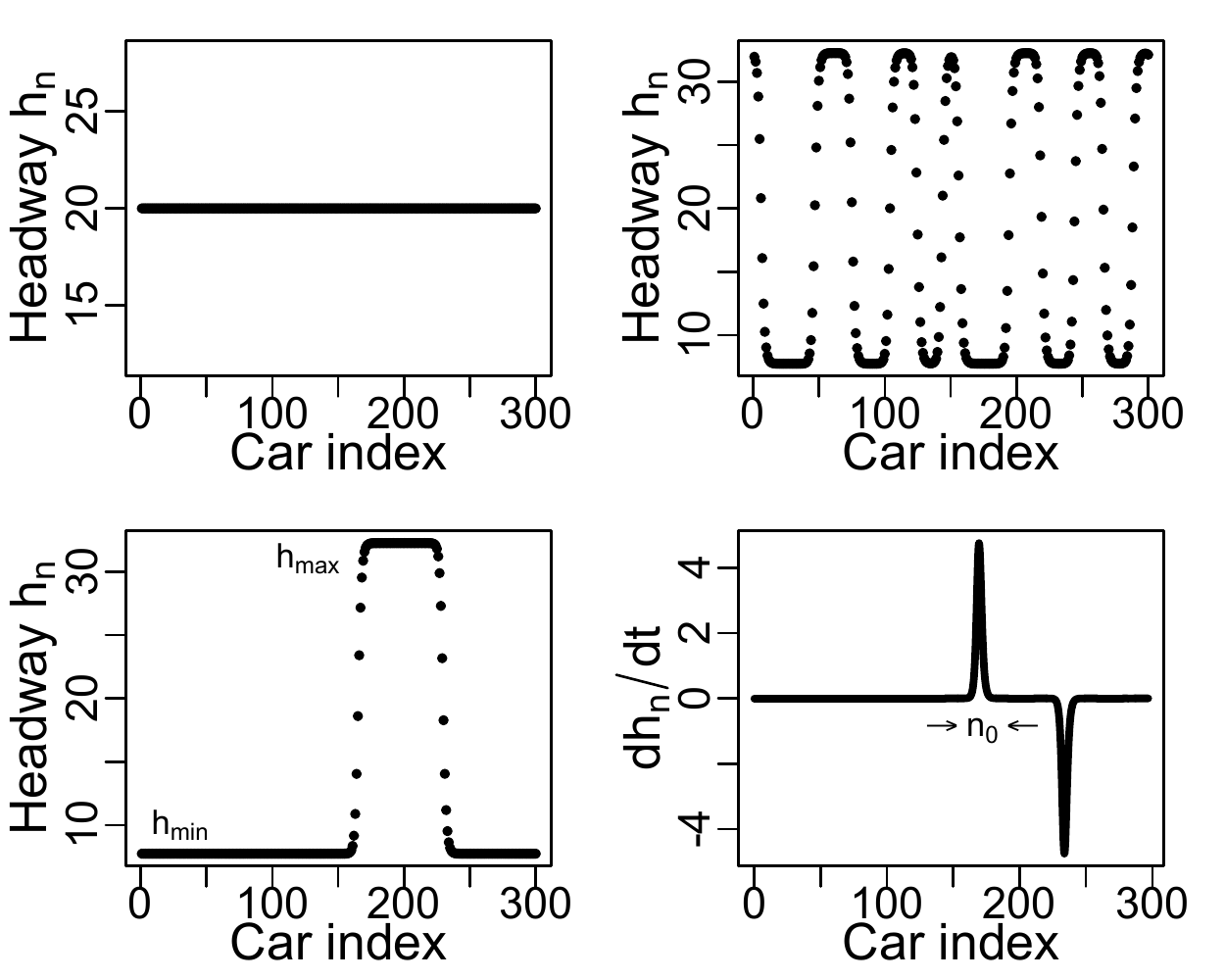}
\caption{Characteristics of solutions to the deterministic GM models. Top left: A uniform headway solution. Bottom left: A single cluster solution, showing clearly the maximum headway and the minimum headway, as well as the transition region in between. Top right: A multi-cluster solution. Bottom right: The time derivative of the headway, showing quasisolitions of opposite charges. The car index axis is made continuous by mapping from time domain into index domain using the velocity of the traveling wave solution, and $n_0$ quantifies the width of the quasisolitons\cite{yangbo}. } 
\label{fig1b}
\end{figure}

It is useful to first understand systematically the nature of simplifications leading to various mathematical models in the literature upon which numerical calculations are performed. For simplicity we assume each driver's action depends only on the state of his own vehicle and the vehicle right in front of him\cite{footnote1}.  The most general form of the car following model along a single lane is given by:
 \begin{eqnarray}\label{acc}
a_n=f_{\{n\}}\left(v_n,v_{n+1},h_n\right)
\end{eqnarray}
where $v_n$ and $v_{n+1}$ are the velocities of the $n^{\text{th}}$ and $(n+1)^{\text{th}}$ car respectively, and $h_n$ is the headway, or the distance between the $n^{\text{th}}$ and $(n+1)^{\text{th}}$ car. The subscript $\{n\}$ indicates the function also depends on various other parameters characterizing the road and weather conditions, driver preferences and emotions, vehicle types, etc. All variables in Eq.(\ref{acc}) are time dependent, though not shown explicitly. In particular, the function $f$ depends on the car index $n$, since each driver is different. The first step of simplification is to assume identical drivers, which can be theoretically justified by replacing $f_{\{n\}}$ with its ensemble average over all vehicles and over time. Thus formally from Eq.(\ref{acc}) we have
\begin{eqnarray}\label{bacc}
a_n=\bar f\left(v_n,v_{n+1},h_n\right), \bar f=\langle f_{\{n\}}\rangle_{\{n\}}
\end{eqnarray}
where the angle bracket represents the ensemble average. In principle even the averaged human driving behavior, as described by $\bar f$, can be quite complicated. The next step of simplification is to include only a few essential features of $\bar f$ in the model, and this has been done at various levels. A number of proposed three-phase microscopic models basically employ highly sophiscated $\bar f$; even with the GM model class\cite{kernercrit,helbing4}, the functional form of $\bar f$ can be quite elaborate. The general optimal velocity (OV) model\cite{bando,JiangR_PRE01,PengGH_PhyA13,GLW_PhyA08} is a subclass of the GM model where $\bar f$ has the simple form 
\begin{eqnarray}\label{ovm}
\bar f\left(v_n,v_{n+1},h_n\right)&=&\kappa_0\left(V_{op}(h_n)-v_n\right)-g\left(\Delta v_n\right)
\end{eqnarray}
where $\Delta v_n=v_{n+1}-v_n$ and $V_{op}(h)$ is the monotonically increasing and bounded optimal velocity function, and $g\left(\Delta v_n\right)$ modifies the acceleration based on the velocity difference between two consecutive vehicles. Given the well-known universal features\cite{xue,nishinari,hasebe,naka} of the cluster solutions to Eq.(\ref{ovm}), we classify the OV models only by parameters extracted from the cluster solution: $h_{\text{max}}$, the maximum headway in the anticlusters, $h_{\text{min}}$, the minimum headway in the clusters, and $n_0$, the intrinsic scale\cite{yangbo} that quantifies the width of the quasisolitons and the strength of interaction between clusters (see Fig.(\ref{fig1b})). Physically, $n_0$ also gives the maximum acceleration in the traffic, and the interaction between clusters is crucial for the time scale of the evolution of the wide moving jams.

While Eq.(\ref{ovm}) only retains a minimal set of the essential features of the realistic $\bar f$ and is deterministic, we will show it is sufficient in capturing many empirical observations with a systematic way of tuning the model. We use the well-studied A-5 North German Highway from Kerner\cite{kernerexp1} as an example. Only the information from the flow-density plot is used, which consists of the free flow part (where the flow depends approximately linearly on the density), the congested part (with a collection of randomly scattering data at higher density with suppresed flow) and the wide moving jam given by the ``J line". The list of statistically robust quantities from the flow-density plot we use are:
\begin{eqnarray}\label{exp}
&&V_{\max}=\lim_{\rho\rightarrow 0}dF/d\rho\sim 42ms^{-1}, \quad F_{\text{max}}\sim 3000 veh/h,\nonumber\\ &&\quad\rho_{\text{cr}}\sim 30 veh/km, \quad F_{dj}\sim 2000 veh/h, \nonumber\\
&&\quad\rho_{dj}\sim 17.5 veh/km, \quad\rho_j\sim 125 veh/km
\end{eqnarray}
Here $\rho_{\text{cr}}$ is the capacity of the highway at which $F_{\text{max}}$, the maximum flow, is observed. $F_{dj}$ and $\rho_{dj}$ are the flow and density downstream of the wide moving jam respectively, while $\rho_h$ is the density within the jam. For the lack of raw traffic data, all the numerical values are rough estimates only, and for our purpose of illustration that is enough, as we do not need to fine-tune the model to simulate the qualitative empirical features. We also assume on average the length of the vehicle $l_c=5m$, and by identifying the parameters of the cluster structure with the characteristic parameters of a wide moving jam we have the following relationship:
\begin{eqnarray}\label{relationship}
&&V_{op}\left(\infty\right)=V_{max}, \quad V_{op}\left(h_{max}\right)=v_{dj}=F_{dj}/\rho_{dj}\nonumber\\
&&V_{op}\left(h_{min}\right)=0, \quad V_{op}\left(h_{cr}\right)=v_{cr}=F_{max}/\rho_{cr}
\end{eqnarray}
The two other characteristic velocities from the flow-density plot are $\widetilde V_J=F_{max}/\left(\rho_{cr}-\rho_j\right)$, the velocity of the downstream front of a wide moving jam, and $\widetilde V_C=\left(F_{max}-F_{dj}\right)/\left(\rho_{cr}-\rho_{dj}\right)$, the velocity of the downstream front between $F_{max}$ and $F_{dj}$. The cluster parameters are given by $h_{max}=\rho_{dj}^{-1}-l_c, h_{min}=\rho_j^{-1}-l_c, h_{cr}=\rho_{cr}^{-1}-l_c$.

We solve Eq.(\ref{relationship}) most simply with a piecewise function passing through $\left(h_{min}, 0\right), \left(h_{cr}, v_{cr}\right), \left(h_{max}, v_{dj}\right)$ and bounded at $V_{max}$, so as to fix the quantitative features of the real traffic dynamics\cite{footnote2}. Defining $h_c=\frac{V_{max}-\widetilde V_C}{v_{cr}-v_{dj}}\left(h_{cr}-h_{max}\right)-l_c$ we have:
\begin{eqnarray}\label{piecewise}
V_{op}\left(h\right)=\left\{
\begin{array}{lr}
0 &  h<h_{min}\\
\frac{v_{cr}}{h_{cr}-h_{min}}\left(h+l_c\right)+\widetilde V_J & h_{cr}>h\ge h_{min}\\
\frac{v_{cr}-v_{dj}}{h_{cr}-h_{max}}\left(h+l_c\right)+\widetilde V_C& h_c>h\ge h_{cr}\\
V_{max} & h\ge h_c
\end{array}
\right.
\end{eqnarray}
 
\begin{figure}
  \centering
  \setbox1=\hbox{\includegraphics[height=4.6cm]{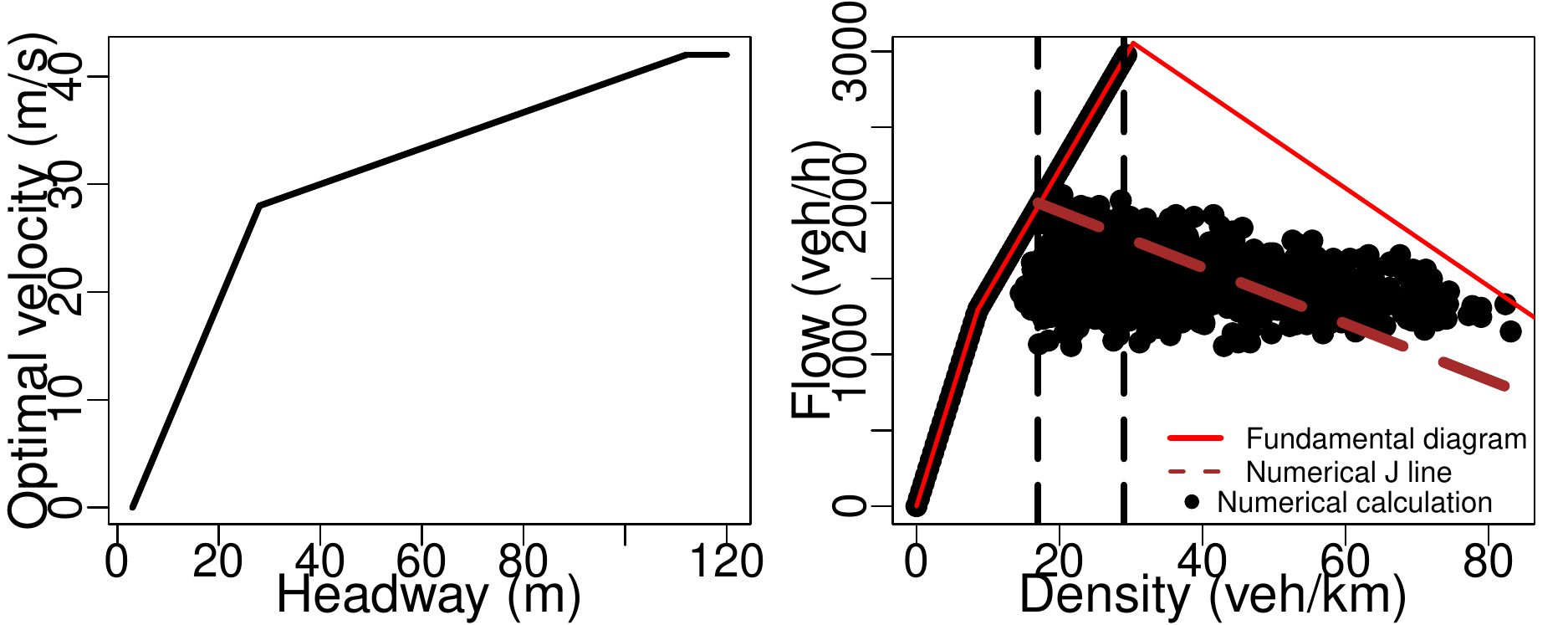}}
  \includegraphics[height=3.5cm]{fig2b.pdf}{\llap{\makebox[\wd1][c]{\raisebox{0.65cm}{\includegraphics[height=1.8cm]{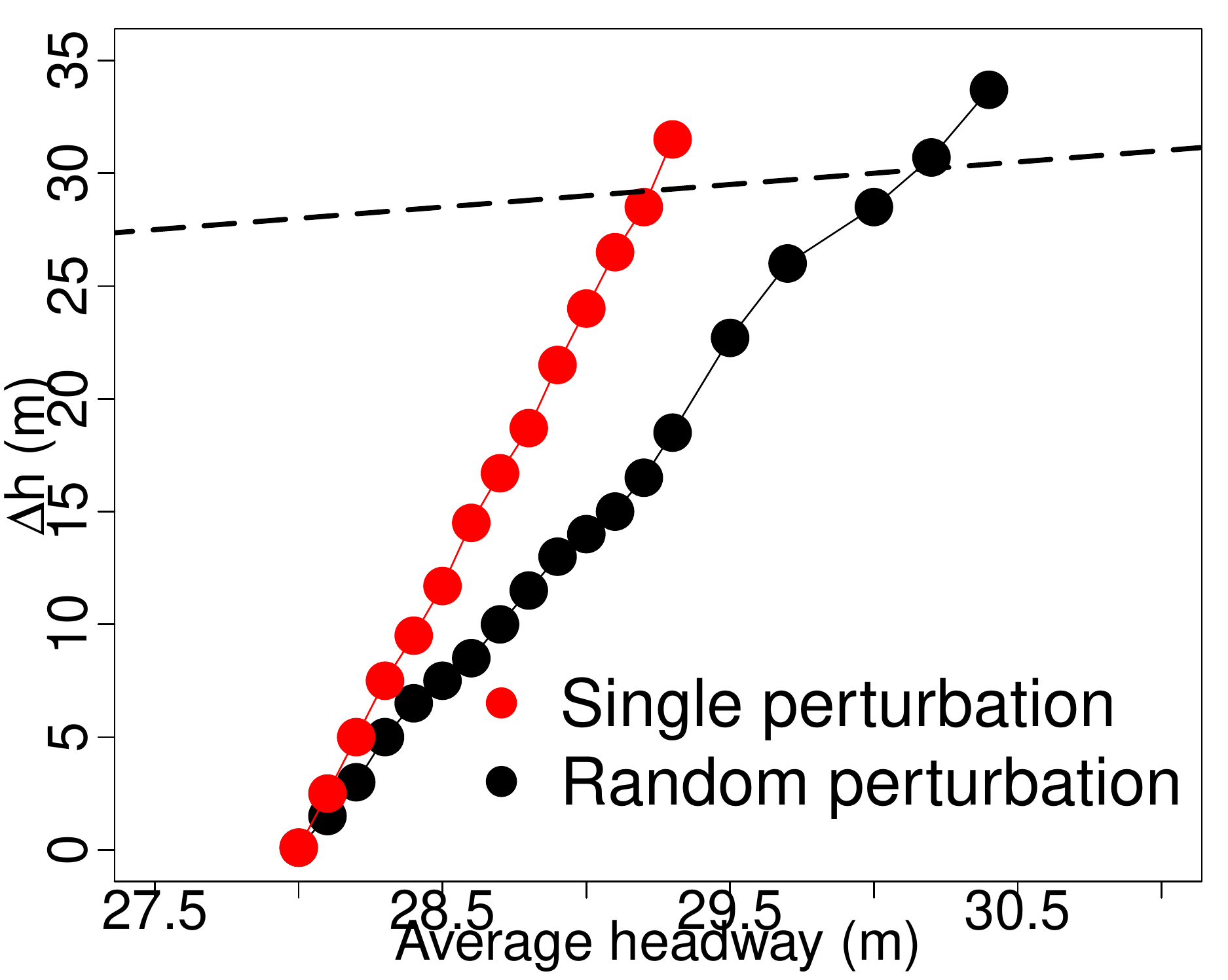}}}}}
  \caption{(Color online) Characteristics of the effective OV model. Left: The piecewise optimal velocity function. Left inset: The magnitude of perturbation ($\Delta h$) needed to form cluster solutions in the metastable region. Cases for the single vehicle perturbation (red) and the random perturbation (black) are plotted. Right: The actual flow-density diagram including both the free flow and the congested flow. The plot is obtained from numerical calcuation with open boundary condition and an on-ramp bottleneck. The left vertical line gives $\rho_{dj}$ and the right vertical line gives $\rho_{cr}$.}
\label{fig2b}
\end{figure}

Thus the fundamental diagram is defined by $V_{op}$ (see Fig.(\ref{fig2b})), capturing the quantitative features of the empirical flow-density plot. The next step is to tune $\kappa_0$ and $g\left(\Delta v_n\right)$ so that the cluster solutions have the desirable $h_{max}, h_{min}$ and $n_0$. Here we adpot the reasonable assumption that the maximum acceleration for the vehicles in the stop-and-go wave should be within the range of $\pm 3ms^{-2}$. The parameters of the cluster solution only determines the qualitative features of the real traffic dynamics, so the fine-tuning of $\kappa_0$ and $g\left(\Delta v_n\right)$ is not important. Given that three parameters need to be fixed, the simplest choice for $g\left(\Delta v_n\right)$ is the AFVD model\cite{GLW_PhyA08} (in the case where $\lambda_2\ne 0$)
\begin{eqnarray}\label{g}
g\left(\Delta v_n\right)=\lambda_1\Delta v_n+\lambda_2|\Delta v_n|
\end{eqnarray}
The effective model is completely defined by Eq.(\ref{bacc})$\sim$Eq.(\ref{exp}) and Eq.(\ref{piecewise})$\sim$Eq.(\ref{g}), with the fitted parameter $\kappa_0=0.1s^{-1}, \lambda_1=6.2$ and $\lambda_2=-2.9$, corresponding to $h_{min}\sim 3m$ and $h_{max}\sim 52m$. 

We now proceed to examine what the effective model predicts about the traffic dynamics. The free flow in the stable phase is given in the region $\rho<\left(h_{max}+l\right)^{-1}\sim 17veh/h$. The metastable region is given by $17veh/km\lesssim\rho\lesssim 30veh/km$. In this region, a large enough perturbation will grow in time and leads to instability of the free flow and formation of jams. The empirical feature that the free flow persists up to the critical density $\rho_{cr}\sim 30 veh/h$ is nicely predicted by the fact that for the metastable region with density smaller than $\sim 30 veh/h$, the perturbation needs to be greater than the average vehicle headway for the free flow to be unstable (see Fig.(\ref{fig2b})), which is unlikely without collisions. Thus the effective model captures $F_{max}$ and $\rho_{cr}$ quite accurately, even though Eq.(\ref{piecewise}) in no way guarantee the stability condition agreeing with the empirical data.

\begin{figure}
\includegraphics[width=8.5cm]{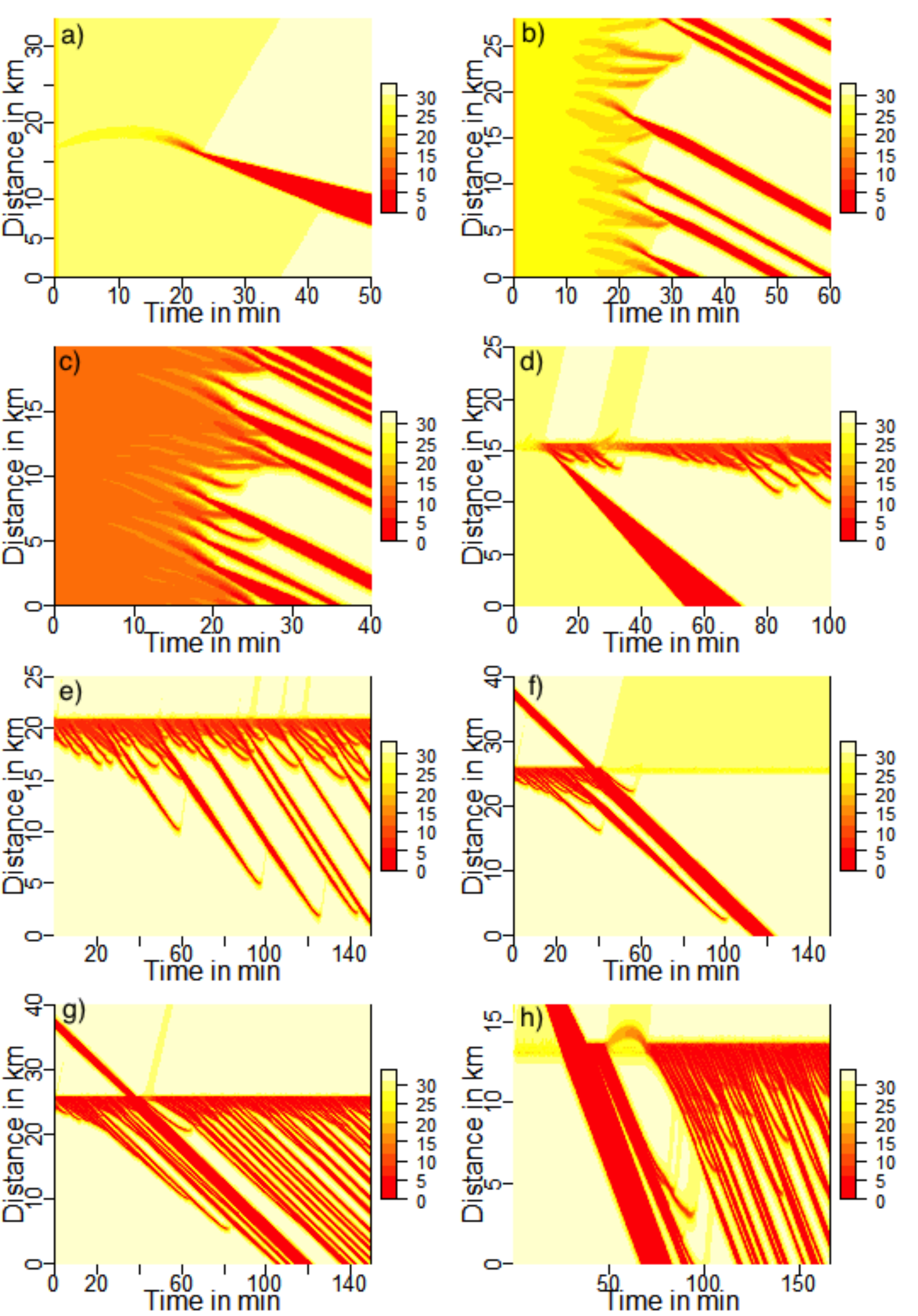}
\caption{(Color online) Various traffic patterns from the effective model with the velocity color plot. Plot a)$\sim$c) are numerical calculations from the periodic boundary condition; the rest are from the open boundary condition with the presence of an on-ramp bottleneck. a). The ``boomerang" behavior with average density close to the coexistence curve, which starts at time =0 min at around distance = 16 km b). The ``pinch effect" in the metastable region. c). Numerous narrow moving jams in the linearly unstable region.  d). A wide moving jam followed by the congested traffic at the bottleneck when the main traffic density is metastable. e). The formation of the congested traffic at the bottleneck when the main traffic density is stable. f). Small $F_{in}$ with metastable main traffic flow, a wide moving jam effectively stops the congested traffic because the density downstream of the jam lower than the that of the main traffic. g). A wide moving jam passes through the botttleneck with its downstream front unaffected. h). A wide moving jam induces congested traffic at the bottleneck. The main traffic density is lower than the density downstream the wide moving jam.} 
\label{fig3b}
\end{figure}

To study the congested traffic and the evolution of the wide moving jams, both periodic boundary condition of a single homogeneous lane and open boundary condition of a single lane with the presence of an on-ramp are simulated. For periodic boundary conditions, the initial condition is chosen to have random fluctuation of the headways with different average density. Though in the long time limit wide moving jams eventually form for average density $\rho\gtrsim 29 veh/km$, the intermediate process can be quite complicated. When the average density is very close to the phase boundary, or the coexistence curve given by $\left(h_{max}+l\right)^{-1}$, very large perturbation is needed to nucleate a wide moving jam via the well-known ``boomerang" behavior\cite{helbing1,kernerbook}; when the average density increases further, the ``pinch effect"\cite{kernerbook} is observed at multiple locations leading to multiple narrow jams; at high density numerous narrow jams form relatively quickly, and over time these narrow jams interact and merge into a few wide moving jams (see Fig.(\ref{fig3b})).

Both the ``pinch effect" and the formation of numerous narrow jams were reported in the literature\cite{kernerbook}, while the existence of the ``boomerang" behavior is still debated\cite{kernerbook,helbing1}. One should note that based on the numerical simulation, it takes $30\sim 60$ minutes for the wide moving jams to eventually emerge from a random initial condition via complex intermediate states. Thus comparing numerical results with fixed number of vehicles and periodic boundary condition to the empirical observations can be extremely tricky. The real world traffic, being an open system, does not maintain its vehicle density and the total number of vehicles over an extended period of time; variation of the average density within the metastable/unstable region leads to a mixture of long lasting intermediate states, numerous narrow jams and occasional wide moving jams. The absence of ``boomerang" behavior may also due to the rarity of very large perturbations on a multi-lane highway, when accidents and bottlenecks are absent.

From both the empirical validation and transportation engineering points of view, numerical simulation of the effective model in the presence of bottlenecks is more crucial. In our simulation with open boundary condition the virtual sensor measures the flow and average velocity of the passing vehicles in \emph{exactly} the same way as the traffic sensors installed in real world highways\cite{helbing2}. We use the idealized initial condition with constant main traffic flow $F_m$ and on-ramp flow $F_{in}$. At low enough $F_{in}$ the free flow is maintained, though in the linearly unstable region the on-ramp flow has to be close to zero. When $F_{in}$ increases, corresponding to the increase in bottleneck strength, the congested flow develops immediately upstream of the bottleneck. This region with length $L_c$  share the characteristics of the ``synchronized flow" (see Fig.(\ref{fig2b}), Fig.(\ref{fig3b})). Narrow jams form upstream of the congested flow, and wide moving jams appear upstream of these narrow jams from merging of the narrow jams and the ``pinch effect". The ``boomerang" behavior is also observed (see Fig.(\ref{fig3b})). The congested flow can be either spontaneous or induced by a passing wide moving jam. In general, $L_c$ can be as long as $4km$ and decreases with the increase of $F_{in}$. 

\begin{figure}
\includegraphics[width=8cm]{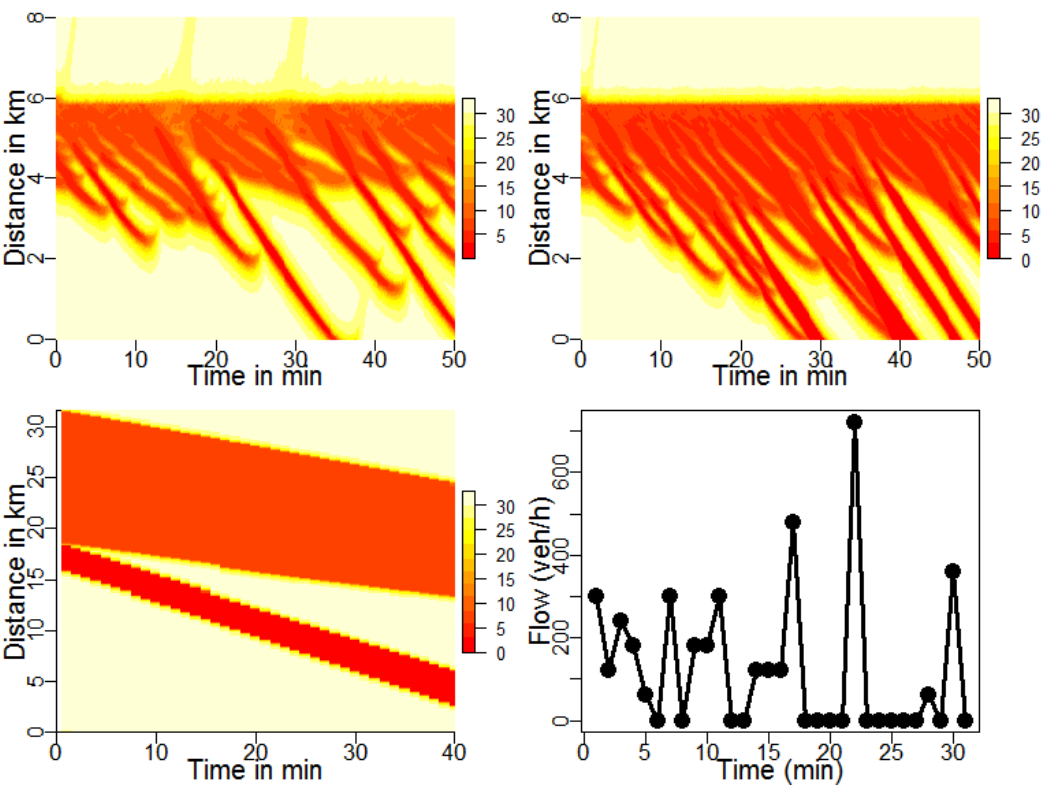}
\caption{(Color online) Additional features predicted by the effective model with the velocity color plot except the bottom right one. Top left: Congested pattern at the bottleneck with $F_m=1800 veh/h$ and $F_{in}=360 veh/h$. Top right: Congested pattern at the bottleneck with $F_m=1800 veh/h$ and $F_{in}=500 veh/h$. Both the narrow jams and wide moving jams emerge closer to the bottleneck, and fluctuating with higher frequencies. Bottom left: Dynamics of a model state with the wide moving jam adjacent to the homogeneous congested state. The velocity of the downstream front of the wide moving jam is unaffected by the traffic conditions downstream of the jam. Bottom right: The flow time series when the traffic density is $80 veh/km$ } 
\label{fig4b}
\end{figure}

Previous GM based models were also criticized\cite{kernerbook} based on some fundamental empirical observations of the congested phase at the bottleneck, as well as on the absence of homogeneous congested traffic (HCT) empirically. In contrast, the effective model we constructed agrees with the empirical observation that increasing the bottleneck strength leads to higher frequency of moving jam emergence and smaller $L_c$ (see Fig.(\ref{fig4b})). In fact this is the most common situation for various different $F_m$. Numerical calculation also shows the metastable phase in the high density region is very narrow. The model actually predicts complicated spatio-temporal structures for congested traffic at very high density, with traffic flow fluctuate between zero to 500 $veh/h$, agreeing with the empirical observation in\cite{kernerbook}. This is simply because small perturbation is linearly unstable even in the region of vehicle density up to $\sim 100 veh/km$.

In conclusion, we show that the physics of many empirical observations of highway traffic dynamics can be captured by a deterministic effective model based on a simple optimal velocity function. In this framework the congested traffic is characterized by long lasting transient states of the model, from which the wide moving jams evolve from the ``pinch effect" or merging of narrow jams. We would like to emphasize though we only used the AFVD model in detail in this work, many different GM models with different parameters can produce cluster solutions with similar $h_{max}, h_{min}$ and $n_0$, thereby giving qualitatively the same agreement with the empirical data. 

Interestingly, our results imply it is probably difficult to distinguish between real traffic dynamics and those from identical autonomous vehicles with simple driving rules based on the empirical data including the flow-density plot and the congestions patterns near the bottlenecks. One should note that even deterministic models can simulate seemingly unpredictable dynamics because the initial condition can be random, which is actually more realistic. In some sense the three-phase models\cite{kimca, kerner1,kerner2,kerner3} could be viewed as extensions of the effective OV models and in principle should be capable of capturing more microscopic-level traffic dynamics. With more advanced data collection and analytics techniques, one can expect to extract more statistically robust quantities to characterize well-defined empirical observations, so that additional levels of sophistication from the modeling point of view can be justified.

\begin{acknowledgements}
This research was partially supported by Singapore A$^{\star}$STAR SERC ``Complex Systems" Research Programme grant 1224504056.  
\end{acknowledgements}


\begin{thebibliography}{99}
\bibitem{footnote1}
Interactions with vehicles further ahead or at the back are generally small and will not change qualitatively the conclusions in this work, see for example \cite{xue}. We can also include the effects of vehicles further away into $\{n\}$ in Eq.(\ref{acc}), without loss of generality.
\bibitem{footnote2}
The most popular form of $V_{op}$ used in the literature is $V_1+V_2\tanh\left(C_1h-C_2\right)$, due to its mathematical simplicity, but not necessarily due to its closeness to the real $\bar f$. While the form gives enough degrees of freedom, it does not have a solution based on the empirical data in Eq.(\ref{exp}). Eq.(\ref{piecewise}) is a simple and straightward way of constructing $V_{op}$ based on any reasonable empirical data.
\end{thebibliography}
\end{document}